\documentclass[12pt]{iopart}
\usepackage{float,graphicx}
\usepackage{epsfig,verbatim}
\newcommand{\noi}{\noindent} 
\newcommand{\be}{\begin{equation}} \newcommand{\ee}{\end{equation}}
\newcommand{\bea}{\begin{eqnarray}} \newcommand{\eea}{\end{eqnarray}}

\begin{document} 

\title{Conformal approach to cylindrical DLA} 
 
\author{A. Taloni$^1$, E. Caglioti$^2$, V. Loreto$^3$ and L. Pietronero$^3$}
 
\address{ 
$^1$Dipartimento di Fisica, Universit\`a di Camerino,
I-62032 Camerino, Italy}
\address{ 
$^2$Universit\`a degli Studi di Roma ``La Sapienza'', Dipartimento di Matematica
P.le A. Moro 5, 00185 Rome, Italy}
\address{ 
$^3$Universit\`a degli Studi di Roma ``La Sapienza'', Dipartimento di Fisica
P.le A. Moro 5, 00185 Rome, Italy and INFM-SMC, Unit\`a di Roma 1
}
\ead{taloni@fisica.unipg.it}

\begin{abstract}

We extend the conformal mapping approach elaborated for the radial
Diffusion Limited Aggregation model (DLA) to the cylindrical
geometry. We introduce in particular a complex function which allows
to grow a cylindrical cluster using as intermediate step a radial
aggregate. The grown aggregate exhibits the same self-affine features
of the original cylindrical DLA. The specific choice of the
transformation allows us to study the relationship between the radial
and the cylindrical geometry.  In particular the cylindrical aggregate
can be seen as a radial aggregate with particles of size increasing with
the radius. On the other hand the radial aggregate can be seen as a
cylindrical aggregate with particles of size decreasing with the
height.  This framework, which shifts the point of view from the
geometry to the size of the particles, can open the way to more
quantitative studies on the relationship between radial and
cylindrical DLA.
\end{abstract}

\maketitle

\section{Introduction}

Since the Diffusion Limited Aggregation (DLA) was introduced in 1981
by Witten and Sander~\cite{ws}, an enormous literature has been
devoted to it. Its paradigmatic role, concerning a variety of pattern
formations in far-from equilibrium processes, such as DBM~\cite{np},
viscous fingering~\cite{pat}, electrodeposition~\cite{4} etc., has
made the DLA one of the most studied models by physicists in the last
twenty years.  Despite the simplicity of its definition, DLA gives
rise to complex branching structures that cannot be described by any
small perturbation of a smooth surface.

DLA was first defined on a two dimensional square lattice. Given a
central particle (seed), new particles are added one by one from a far
away region. The new particle performs a random walk and, when it
touches the aggregate, it becomes part of it. This process is repeated
as many times as the number of particles composing the cluster.

\noindent One can also define the growth process in a cylindrical
geometry~\cite{Meakinfirst}. The initial seed in this case is a base line
and particles 
are released from a far away line parallel to the base line.  Since in
practice the length of the base line is finite and one uses periodic
boundary conditions, topologically the growth occurs on the surface of
a cylinder. We will refer to this model as ``cylindrical DLA''.

The relationship between radial and cylindrical aggregates constitutes
a major puzzle for the theorists since the fractal and multifractal
properties of the aggregate seem definitely to depend (though in a
weak way) on the geometry where the growth process occurs. Indeed,
whereas the two processes (radial and cylindrical) give rise to
basically similar structures, there are small but robust differences
that persist to the asymptotic limit ~\cite{12}, and it is not obvious
to conclude that they are just due to finite size effects. The value
of the fractal dimension of radial DLA still represents an open and
controversial question ~\cite{5b,5bb,mandel-boz,ball2,menshutin},
whereas the most accredited one sets approximately  around $1.71$. On the
other hand, the fractal dimension of the cylindrical DLA seems to be
not so sensitive to different measurements techniques and its value,
measured by box-counting method, has been approximated by
1.65~\cite{5bb,evertsz}. Nevertheless, it is interesting to note that
the value of fractal 
dimension measured for a circular crosscut in radial aggregates is
equal to the one obtained for the intersection set of cylindrical DLA
($D_I\sim0,65$)~\cite{10,11,12}.  Finally,  it should be stressed that
recent off-lattice
simulations using hierarchical maps algorithm propose an equivalence
between fractal dimension of radial and cylindrical
DLA~\cite{ball2}. 

The problem of the correct definition and computation of the fractal
dimension is closely linked to the matter of self-similarity. For
radial clusters the overall form of aggregates slowly changes during
the growth, exhibiting a multi-armed shape and progressively filling
the space more uniformly~\cite{8}. Indeed the structures generated by
simple self-similar scaling models suggest that the lacunarity
decreases with increasing size~\cite{6,7,8}. On the other hand
cylindrical DLA does not present deviations from self-similarity,
consequently his lacunarity is constant as it grows up. Roughly
speaking, these discrepancies are probably due to the fact that in
cylindrical DLA the cylinder size is fixed and it does not depend on
the growth process itself~\cite{5bb}, whereas the ratio between
cut-offs (size of particles and size of cluster) in radial geometry is
constantly changing. In this sense the cylinder geometry offers a
conceptual advantage for a theoretical discussion~\cite{5bb,19},
because it defines a unique growth direction and it allows to vary
independently the size of the base line and the height, which are
instead intrinsically linked in the radial geometry.

The two aggregates differ also on the initial \emph{non-equilibrium}
stages of their growth. Although the self affine scaling regime of the
cylindrical DLA constitutes a well-known
topic~\cite{meakin,evertsz,vannimenus} lying on the surfaces' growth phenomena
framework, in the radial geometry the matter of the existence of two
different length scales is  an open  controversial
question~\cite{radial2scaling,25}.

Recently an elegant representation of radial DLA growth in terms of
iterated conformal maps has been introduced by Hastings and
Levitov~\cite{13,14}. This formulation makes available the powerful
tools of analytic function theory, leading to accurate measurements in
multifractal properties of the aggregates~\cite{15,16} and important
theoretical works on the structures of these ones~\cite{24,18}.

The aim of this paper is to extend to the cylindrical geometry the
analytic procedure of the conformal mapping elaborated for the radial
case. The main idea behind our conformal approach is to map the
unitary circle onto the interface of cylindrical cluster passing
through the radial geometry. We introduce, in particular, a complex
function that conformally maps the exterior of the unit circle onto
the interior of an infinite stripe. Such a function allows us to shift
from the radial geometry to the cylindrical one and \emph{vice
versa}. The composition of such a function with an Hastings and
Levitov-like function would lead to an analytic map that transforms
the exterior of the unit circle onto the complement of a cluster,
growing in a stripe with periodic boundary conditions. In this way we
can study the growth process of a cylindrical cluster as well as that
of its radial deformation: the dimension is the same for both the
aggregates as they are related by isomorphism. The same
consideration also applies to the case of a ``real'' radial DLA that
can be deformed onto a cluster growing in a periodic stripe, by
composing the functions as before. In this framework the question of
the relationship between the radial and the cylindrical DLA appears to
be a natural and well-defined problem. In particular the cylindrical
aggregate can be seen as a radial aggregate with particles of size
increasing with the radius. On the other hand the radial aggregate can
be seen as a cylindrical aggregate with particles of size decreasing
with the height.  This framework, which shifts the point of view from
the geometry to the size of the particles, can open the way to more
quantitative studies on the relationship between radial and
cylindrical DLA.

The outline of the paper is as follows. In section II we recall the
conformal mapping model for the radial DLA and we introduce the
complex function that transforms the exterior of the unit circle onto
an infinite periodic strip. Moreover we present the conformal mapping
rules that allow to build up the cylindrical cluster. In section III
we focus on the problem of the so-called unphysical particles and we
discuss a new procedure to discard them. In section IV we will deal
with the question of the dimension of cylindrical DLA.  We discuss the
scaling behavior of
the overall height of a cylindrical cluster as a function of a
universal scaling variable. The self-affine and self-similar behaviours of
the overall cluster's height are  then carefully examined.  

\section{Conformal mapping approach}

\subsection{Radial DLA}

We briefly recall the conformal mapping formulation of radial
DLA~\cite{13}.

\noindent Let us consider an analytic function that maps the unit
circle in the mathematical $w$-plane onto the complement of the
cluster of $n-1$ particles in the physical $z$-plane. Such a function
is derived from the composition of elementary maps
$\varphi_{\lambda,\vartheta}$:

\begin{equation}
\Phi^n(w)=\Phi^{(n-1)}(\varphi_{\lambda_n,\vartheta_n}(w)).
\label{1}
\end{equation}

\noindent The conformal transformation $\varphi_{\lambda,\vartheta}$
maps the exterior of the unit circle to the exterior of the unit
circle with a bump of linear scale $\sqrt{\lambda}$ around the point
$e^{i\vartheta}$. A choice of this function that is free of global
distortion is~\cite{20}

\begin{eqnarray}
\varphi_{\lambda,0}(w)=w^{\frac{1}{2}}\left\{\frac{(1+\lambda)}{2w}(1+w)\times
\right.\nonumber\\ \left. \left[1+w+w\left(1+\frac{1}{w^2} -
\frac{2}{w}\frac{1-\lambda}{1+\lambda}\right)^{\frac{1}{2}}\right]-1\right\}^{\frac{1}{2}},
\label{fi0}
\end{eqnarray}

\noi with 

\begin{equation}
\varphi_{\lambda,\vartheta}(w)=e^{i\vartheta}\varphi_{\lambda,0}(e^{-i\vartheta}w).
\label{fi}
\end{equation}

\noindent In order to obtain in the physical $z$-plain particles of
fixed size $\sqrt{\lambda_0}$ the size of the $n-$th bump
$\sqrt{\lambda_n}$ has to be

\begin{equation}
\sqrt{\lambda_n}=\frac{\sqrt{\lambda_0}}{|(\Phi^{n-1})'(e^{i\vartheta_n})|}.
\label{2}
\end{equation}

\noindent We emphasize that this approximation is valid only to the
first order. Fluctuation of the magnification factor
$|(\Phi^{n-1})'(e^{i\vartheta_n})|$ over the unit circle on the scale
of $\sqrt{\lambda_n}$, can generate particles of very unequal sizes
~\cite{20}~\cite{21}(see also Section III).
\noindent Furthermore it is immediate to prove that the harmonic
probability on the boundary of a real cluster in $z$ translates to an
uniform measure on the unit circle:

\begin{equation}
P(s,ds)=d\vartheta,
\label{2bis}
\end{equation}

\noindent where $z(s)$ is a point on the cluster's interface, and $ds$
is an infinitesimal arc centered on this point.

In the first papers~\cite{13,20} it was also stressed that the scaling
of the cluster radius, $R_n$, at large $n$ is well characterized by
the scaling of the first Laurent coefficient of the mapping
function~(\ref{1}). Indeed the Laurent expansion of $\Phi^{n}$ is

\begin{equation} 
\Phi^{n}(w)=F_1^{(n)}w+F_0^{(n)}+\sum_{k=1}^{+\infty}F_{-k}^{(n)}w^{-k}
\label{laufi},   
\end{equation}

\noindent so that $\Phi^{n}(w)\sim F_1^{(n)}w$ as $w\to\infty$. Since
one expects for the radius $R_n$ a scaling form
$R_n\sim(n)^{\frac{1}{D}}\sqrt{\lambda_0}$, where $D$ is is the
fractal dimension of the cluster, we can assume

\begin{equation}
F_1^{(n)}\sim(n)^{\frac{1}{D}}\sqrt{\lambda_0}
\label{3}.   
\end{equation}

\noindent The analytical form of $F_1^{(n)}$ is well-known~\cite{20}

\begin{equation}
F_1^{(n)}=\prod_{k=1}^{n}\left(1+\lambda_k\right)^{\frac{1}{2}}
\label{4}.   
\end{equation}

\noindent The scaling law (\ref{3}) offers a very convenient way to
measure the fractal dimension of the growing cluster~\cite{20};
moreover, assuming that the total area of the cluster scales
as $A_n\sim n\sqrt{\lambda_0}$~\cite{20,21}, one can write
the relation (\ref{3}) as~\cite{18,21,22}:

\begin{equation}
R_n\equiv F_1^{(n)}\sim A_n^{\frac{1}{D}}.
\label{5}   
\end{equation}

\noindent Clearly this is true provided that individual particles
areas have a sufficiently narrow distribution~\cite{21}
(see Section III).

\subsection{Cylindrical DLA}

The essential ingredient of our conformal approach for the cylindrical
DLA is the modification of the relation (\ref{2}) which gives rise to
a deformed radial cluster with particles of unequal sizes. That
cluster will after turn into a cylindrical DLA by mean of a suitable
function that shifts from the radial geometry to the cylindrical one.
In this way we can adapt the conformal theory developed for the radial
case to the cylindrical case, without changing the elementary function
(\ref{fi0}), as instead proposed in~\cite{ball2}.

In the radial representation, the conformal mapping (\ref{1}) can be
expressed as

\be \Phi^n(w)=\Phi^{(0)} \circ\varphi_{\lambda_1,\vartheta_1}
\circ\varphi_{\lambda_2,\vartheta_2}
\circ\varphi_{\lambda_3\vartheta_3}\circ\cdots\circ\varphi_{\lambda_n,\vartheta_n}(w).
\label{cazzo}
\ee

\noindent As originally mentioned in~\cite{20} the choice of the
initial map $\Phi^{(0)}(w)$ is flexible and one can expect the
asymptotic shape of the cluster to be independent from this
choice. The simplest choice of $\Phi^{(0)}(w)$ was
$\Phi^{(0)}(w)=w$~\cite{13,20}, which turns the previous definition of
the mapping function in:

\begin{equation}
\Phi^n(w) = \varphi_{\lambda_1,\vartheta_1}
\circ\varphi_{\lambda_2,\vartheta_2}
\circ\varphi_{\lambda_3\vartheta_3}\circ\cdots\circ\varphi_{\lambda_n,\vartheta_n}(w).
\label{6}
\end{equation}

\begin{figure}[!t]
\begin{center}
\includegraphics[width=8cm]{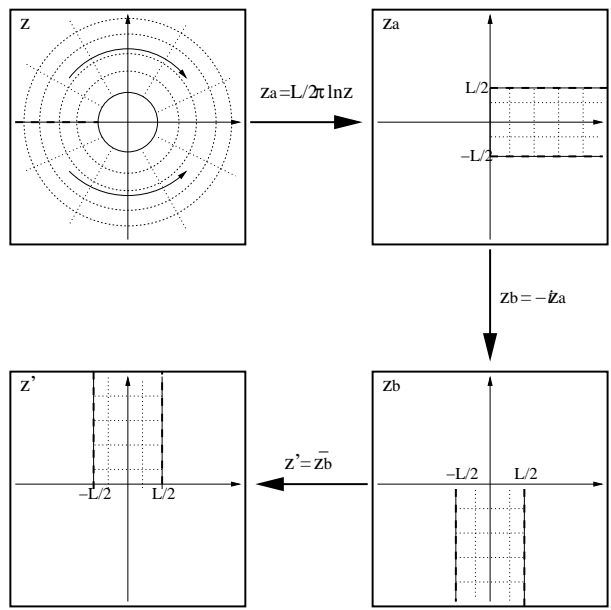}
\caption{The action of (\ref{8}) on the complex $z$-plane is
  schematically represented here. We divide the function in three
  steps. First box($z\to z_a$): after a cut along the negative part of
  the real axis (dashed line), the two parts on the half-plane $\Re
  e(z)<0$ ($\Im m(z)>0$ and $\Im m(z)<0$) are rotated in the sense of
  the arrows, just as a Chinese-fan-like closure. Second box($z_a\to
  z_b$): the closed Chinese fan is rotated by an angle
  $=-\frac{\pi}{2}$. Third box($z_b\to z'$): the fan is rotated with
  respect the real $z_b$ axis.  The closure of the Chinese-fan is the
  stripe on the $z'$ complex plane where the growth of a cylindrical
  aggregate occurs.}
\label{fig1}
\end{center}
\end{figure}

Let us now consider the function

\be
\Xi^L_0(z)=\overline{-i\frac{L}{2\pi}\ln(z)},
\label{7}
\ee

\noi which maps the exterior of the unit circle in the $z$-plain, into
the interior of a infinite stripe with periodic boundary conditions in
the half-plain $\Im m( z')>0$. $L$ is a real parameter whose physical
meaning is the width of the stripe in the $z'$-plane.  A similar
function was introduced for the first time in~\cite{18} with regard
to the growth in a channel. If we call the polar coordinates of the
$z$-plain $\rho$ and $\theta$ we rewrite (\ref{7}) as:

\be
\Xi^L_0(z)=
\left\{
\begin{array}{ccc}
x(\rho,\theta) & = &\frac{L}{2\pi}\theta\\
y(\rho,\theta) & = &\frac{L}{2\pi}\ln\rho
\end{array}
\right.
\label{8}
\ee

\begin{figure}[!t]
\begin{center}
\includegraphics[width=8cm]{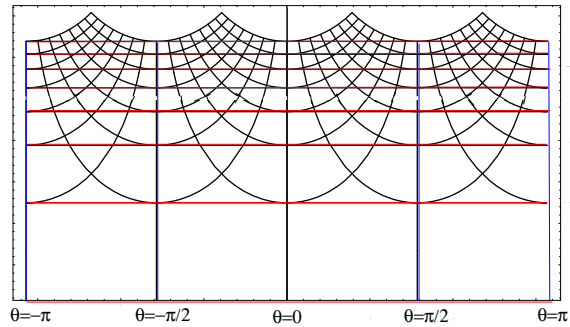}
\caption{Deformation in $z'$ of a square lattice in $z$ complex plane.}
\label{fig2}
\end{center}
\end{figure}

\noi where $x$ and $y$ are respectively the real and the imaginary
part of $z'$. It is possible to imagine the action of (\ref{7}) on the
$z$-plane as the closure of a Chinese fan (See Fig.\ref{fig1}). For
sake of clarity Fig.[\ref{fig2}] shows as a square lattice on the
radial plane $z$ is deformed in the $z'$ plane by (\ref{8}).

\begin{figure}[!t]
\begin{center}
\includegraphics[width=8cm]{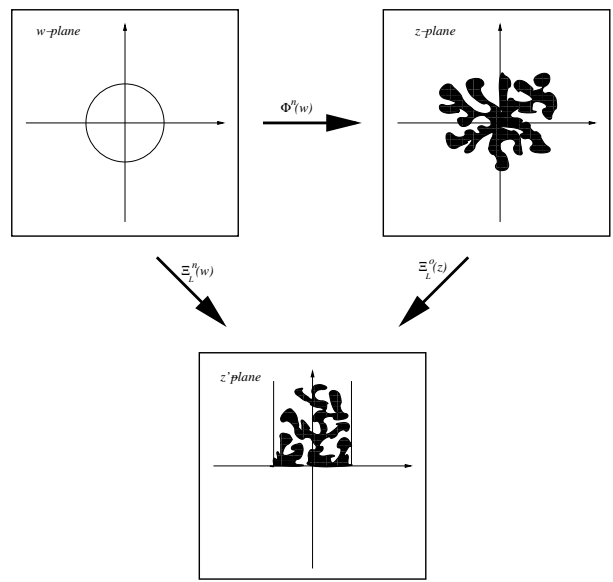}
\caption{Sketch of the conformal approach used to construct a
  cylindrical aggregate. First we map the unit circle onto the
  interface of a radial cluster in the $z$-plane, by the usual
  Hastings and Levitov formula (\ref{1}). After, we compose this
  function with (\ref{7}). This consists in a mapping from unit circle
  to the cluster growing on the $z'$-plane.}
\label{fig3}
\end{center}
\end{figure}

\noi Our choice of the function (\ref{7}) seems to be the \emph{natural link}
between radial and cylindrical geometry. Indeed if we compose
(\ref{7}) with (\ref{6}) we obtain a conformal transformation that
maps the unit circle, in the $w$-plane, to the interface of a
cylindrical aggregate in the $z'$-plane passing through its ``radial
version'' in the $z$-plane (See Fig.\ref{fig3}):

\be  
\Xi^L_n(w)=\Xi^L_0\circ\Phi^n(w)
\label{9}.
\ee 

\noi In fact it is possible to look at the (\ref{9}) as a particular form
of the (\ref{cazzo}) with the $\Phi^{(0)}$ replaced by $\Xi^L_0$.

\begin{figure}[!t]
\begin{center}
\includegraphics[width=8cm]{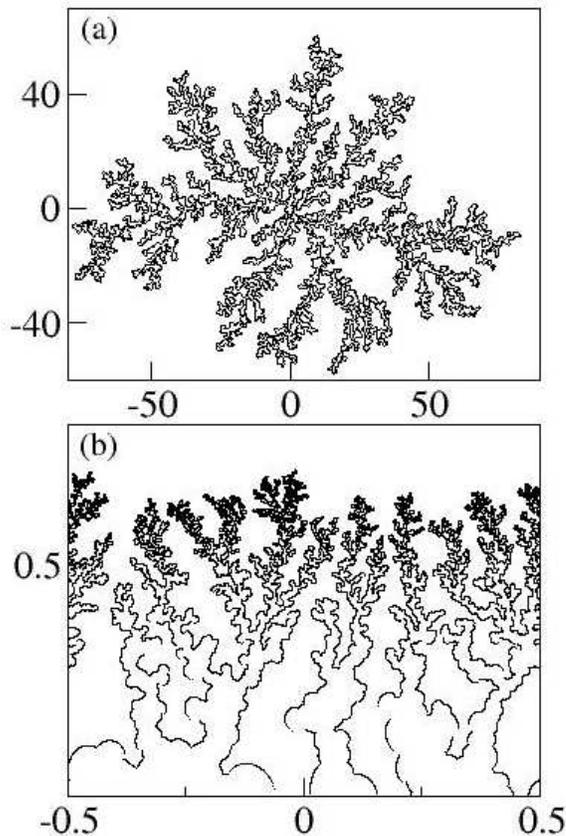}
\caption{Radial DLA: (a) radial DLA grown with the rules (\ref{1}), (\ref{2})
  and the acceptance criterion expressed in Section III. (b)
  Cylindrical deformation of the same cluster,  obtained by
  composition of (\ref{1}) with (\ref{7}), note as in this plane the
  particle sizes decrease following (\ref{10}). The simulation was
  performed with 10000 particles, $\lambda_0=0.1$ and cylinder size
  $L=1$. }
\label{fig4}
\end{center}
\end{figure}

\begin{figure}[!t]
\begin{center}
\includegraphics[width=8cm]{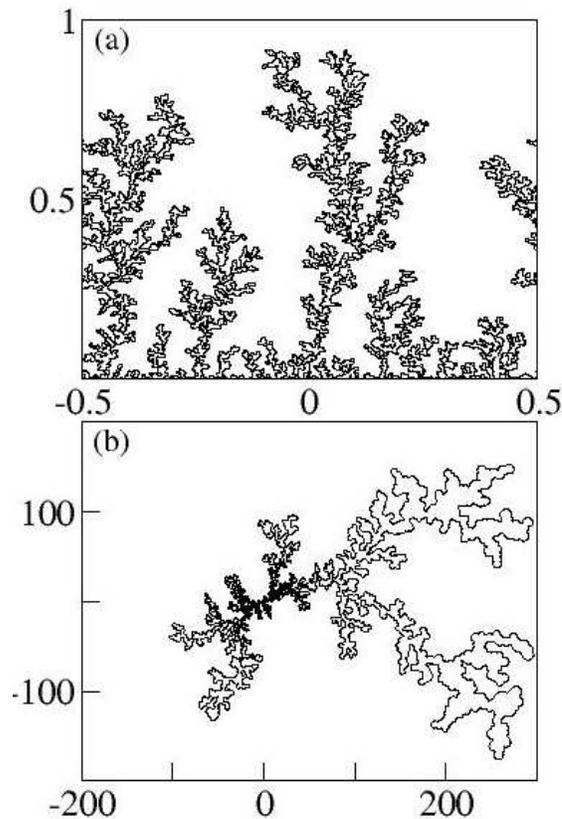}
\caption{Cylindrical DLA:  (a) cylindrical DLA  grown with the rules (\ref{9}),
  (\ref{11}) and the acceptance criterion expressed in Section
  III. (b) Radial aggregate  grown with (\ref{1}), (\ref{13}),
  conformal deformation  of  DLA showed in panel (a) using (\ref{7}). Here
  $n=10000$, $\lambda_0=5\times 10^{-6}$ and $L=1$.}
\label{fig5}
\end{center}
\end{figure}

Let us now look, with the help of Figures [\ref{fig4}] and
[\ref{fig5}], at the effect of the different transformations
introduced.  If we grow a cluster according to (\ref{1}), (\ref{2}),
we have the ``true'' radial aggregate in the physical growth space
($z$) (Fig.\ref{fig4}(a)). On the other hand it is possible to see
this cluster deformed by (\ref{7}) in the $z'$-plane
(Fig.\ref{fig4}(b)). Consequently in $z'$ the size of the
particles ($\sqrt{\lambda_0}$ in $z$) become smaller and smaller as
the cluster size increases:

\be
\sqrt{\lambda_n}^{(z')}=
\frac{\sqrt{\lambda_0}L}{2\pi\left|\Phi^{(n-1)}(e^{i\vartheta_n})\right|}=\frac{\sqrt{\lambda_0}L}{2\pi e^{\frac{2\pi}{L}\Im m
[\Xi^{(n-1)}(e^{i\vartheta_n})]}}
\label{10}
\ee

\noi On the other hand if we want to construct a \emph{real}
cylindrical aggregate in $z'$, we have to modify (\ref{2}) in order to
obtain particles of fixed size $\sqrt{\lambda_0}$ in this plane. To
this end we have to divide the characteristic scale by the Jacobian of
the mapping (\ref{9}):

\begin{eqnarray}
\sqrt{\lambda_n} &=&
\frac{\sqrt{\lambda_0}}{\left|\Xi^{(n-1)'}(e^{i\vartheta_n})\right|}
 =  \frac{2\pi \sqrt{\lambda_0}\left|\Phi^{(n-1)}(e^{i\vartheta_n}) \right|}
{L\left|\Phi^{(n-1)'}(e^{i\vartheta_n})\right|}= \nonumber\\
                 &=&\frac{2 \pi\sqrt{\lambda_0} e^{\frac{2\pi}{L}\Im m[\Xi^{(n-1)}(e^{i\vartheta_n})]}}
{L\left|\Phi^{(n-1)'}(e^{i\vartheta_n})\right|}.
\label{11}
\end{eqnarray}

\noi Therefore we can grow a cylindrical cluster in the physical plane
$z'$ using (\ref{9}) and (\ref{11}) (Fig.\ref{fig5}(a)) just as
we grew a radial cluster in $z$-plane using (\ref{1}) and
(\ref{2}). In analogy with what done for the radial case, it is
possible to see the cylindrical cluster deformed by the inverse of
(\ref{7}) in the $z$-plane. In this case the size of the particles is
increasing exponentially with the height ($y$ in (\ref{8})) of the
aggregate (Fig.\ref{fig5}(b)):

\be
\sqrt{\lambda_n}^{(z)}=\frac{2
\pi\sqrt{\lambda_0}\left|\Phi^{(n-1)}(e^{i\vartheta_n})\right|}{L}=
\frac{2\pi\sqrt{\lambda_0}e^{\frac{2\pi}{L}\Im
m[\Xi^{(n-1)}(e^{i\vartheta_n})]}}{L}
\label{12}
\ee

\noi The complex potential on the stripe is given by

\be
\Upsilon^{(n)}(\overline{z'})=\ln[(\Xi^{(n)})^{-1}(\overline{z'})],
\label{potential}
\ee

\noi so it is quickly verified that $P(s,ds)=d\vartheta$ on the unit
circle as in radial case. Notice that the boundary conditions of the
Laplacian field $\nabla P=\frac {\Phi^{(n)'}}{|\Phi^{(n)}|}$ at
infinity will be automatically changed from $\nabla
P\sim\frac{\hat{r}}{r} $ to $\nabla P\sim cost \, \hat{y} $~\cite{18}.

\section{Construction of the aggregate}

On the original work on the radial DLA~\cite{13}, it was assumed 
the rule (\ref{2}) was sufficient to produce particles with nearly
equal areas ($\sim \lambda_0$). However, as we have noticed in the
discussion after (\ref{2}), this is not true in general: large
particles tend to appear within fjords and seal completely large
otherwise deeply invaginated regions, where the magnification factor
$\Phi '$ is not constant around the bump of (\ref{fi}).  Since our
cylindrical growth model is sensitive to this problem much more than
the radial one, we introduce here a fast and efficient method to
eliminate too large particles.

For radial clusters, a method for the problem of abnormally stretched
particles was proposed in~\cite{20} by choosing an optimal shape of
bump produced by the elementary mapping
$\left|\varphi_{\lambda,\vartheta}\right|$;  
more recently another sophisticated technique has been introduced by
Stepanov and Levitov~\cite{21}. This method consists in evaluating the
particle areas with unprecedented accuracy and discard the particles
whose surface exceeds an acceptance threshold.  \noi Although the
problem of the unphysical particles seems to be crucial in the
evaluation of the cluster dimension by means of (\ref{5}), for radial
clusters numerical simulations suggest that the presence of large
particles is irrelevant for the cluster size scaling~\cite{20,21}.

In the cylindrical case the situation is more subtle. We start by
noticing that we can write (\ref{11}) as:

\be
\sqrt{\lambda_n}=\frac{\sqrt{\lambda_n}^{(z)}}{\left|\Phi^{(n-1)'}(e^{i\vartheta_n})\right|}.
\label{13}
\ee

\begin{figure}[!t]
\begin{center}
\includegraphics[width=8cm]{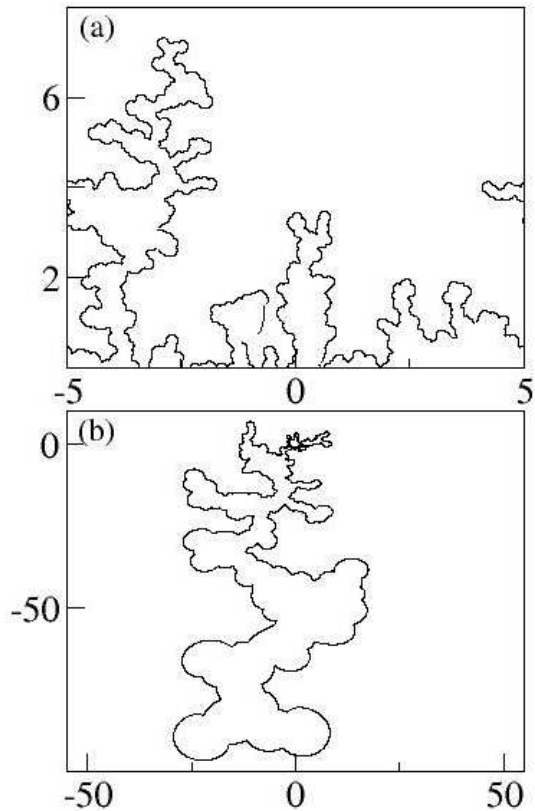}
\caption{Cylindrical DLA: (a) cylindrical aggregate  obtained by use of
  (\ref{9}), (\ref{11}) and its radial \emph{alter ego} (b). It is
  possible to see 
  how the appearance of abnormally stretched particles is frequent in
  this models. Other parameter simulation are: $n=500$, $L=10$ and
  $\lambda_0=0.011$} 
\label{fig6}
\end{center}
\end{figure}

\noi The comparison of (\ref{13}) with (\ref{2}) clearly shows that it
is possible to look at the aggregate grown according to (\ref{9}) and
(\ref{11}) as a radial cluster with particles whose mean size
constantly increases with the (\ref{12}). Since for large $n$ the size
of the particles composing the radial version of the cylindrical DLA
(see Fig.\ref{fig6}(a)) is much larger than the one in the
original radial model
(i.e. $\sqrt{\lambda_n}^{(z)}\gg\sqrt{\lambda_0}$ definitely), the
scale $\sqrt{\lambda_n}$ in (\ref{13}) will be likely larger than the
one in (\ref{2}). Consequently we can expect a more important presence
of filling particles than in the \emph{simple} radial model (\ref{1}),
(\ref{2}). An example of a typical cylindrical cluster
(Fig.\ref{fig6}(a,b)) demonstrates that
there is a high number of particles that macroscopically affect the growth.

\noi Our method to filter abnormally large particles is as follows. We
recognize that every time we grow a semicircular bump with (\ref{fi}),
we generate two new branch-cuts in the map $\Phi^{(n)}$. Each
branch-cut has a pre-image on the unit circle that can be easily
derived from (\ref{fi}):

\be
w^{\pm}=e^{i\vartheta^{\pm}}=\frac{(1-\lambda)\pm
2i\sqrt{\lambda}}{(1+\lambda)}.
\label{14}
\ee

\noi These points can be labeled by two indices~\cite{23}:
$w_{j,n}$. The index $j$ represents the generation when the branch-cut
was created (i.e. when the $j$th particle was grown). The index $n$
stands for the generation at which the analysis is being done
(i.e. when the cluster has $n$ particles). Indeed, after each
iteration, the pre-image of each branch-cut moves on the unit circle
but its physical positions does not change, so that we have a
different list of ``exposed'' branch-cut pre-images $\{w_{j,n}\}$ each
time we grow a particle. Indeed suppose that the list $\{w_{j,n-1}\}$
is available. In the $n$th generation we grow a new bump around the
angle $\vartheta_n$ whose brunch-cuts are~\cite{20}:

\be e^{i\beta^{\pm}}=\varphi_{\lambda_n,\vartheta_n}(w^{\pm}_{n,n}).
\label{15}
\ee

\noi If one or more of the branch-cut pre-images in the updated list
$\{w_{j,n-1}\}$ is covered by the $n$th particle (i.e. $w_{j,n-1}\in
[\beta^+,\beta^-]$ for some $j$), it will not be registered on the new
list $\{w_{j,n}\}$, otherwise it will be included as

\be
 w_{j,n}=\varphi_{\lambda_n,\vartheta_n}^{-1}(w_{j,n-1}).
\label{16}
\ee

\noi These values, obtained by varying $j$, and the sorted new pair
$w^{\pm}_{n,n}$, will compose the $n$th list. The analytical form of
$\varphi_{\lambda,0}^{-1}$ is~\cite{20}:

\be  
\varphi_{\lambda,0}^{-1}=\frac{\lambda
w^2\pm\sqrt{\lambda^2w^4-w^2[1-(1+\lambda)w^2][w^2-(1+\lambda)]}}{1-(1+\lambda)w^2}
\label{17}
\ee

\noi and the inverse mapping $\varphi_{\lambda,\vartheta}^{-1}$ is
given by
$\varphi_{\lambda,\vartheta}^{-1}(w)=e^{i\vartheta}\varphi_{\lambda,0}^{-1}(e^{-i\vartheta}w)$.
Notice that the inverse function $\varphi_{\lambda,\vartheta}^{-1}$ is
analytic on the unit circle only outside the arc $[\beta^+,\beta^-]$.

\begin{figure}[!t]
\begin{center}
\includegraphics[width=8cm]{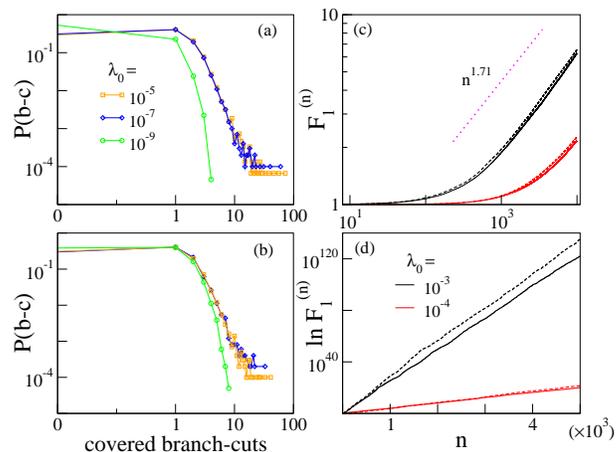}
\caption{Controlled growth of the aggregate: (a) covered branch-cut
  distributions (P(b-c)) for radial model (\ref{1}) (\ref{2}), and (b)
  for cylindrical model (\ref{9}) (\ref{11}), obtained for different
  values of the particle size $\sqrt{\lambda_0}$; statistics was taken
  over 20 different realizations of the firsts 1000 growth steps
  setting $L=1$. (c) scaling behavior of $F^{(n)}_1$ for radial DLA as
  predicted by (\ref{5}) for uncontrolled growth model (solid lines)
  and for the proposed one (dashed lines), axes are set in logarithmic
  scale; the expected power-law growth of radius was drawn for
  reader's convenience (dotted curve). (d) Scaling behavior of the
  overall height of cylindrical cluster (see (\ref{18}) and
  (\ref{19})): it is apparent as the controlled and uncontrolled
  growth models coincide in the limit $\frac{\sqrt{\lambda_0}}{L}\to
  0$, here $L=10$.}
\label{fig7}
\end{center}
\end{figure}

The branch-cut covered distributions for radial model (\ref{1})
(\ref{2}) and cylindrical model (\ref{9}) (\ref{11}) are displayed in
panel (a) and (b) of  Fig.\ref{fig7} respectively. The two
distributions are equally peaked about one covered branch-cut and seem
to show similar shape and the same power-law decreasing trend on the
tails. Furthermore it is apparent that such tails tend to become less
pronounced as the particle size decreases and to completely disappear
in the limit $\sqrt{\lambda_o}\to 0,\frac{\sqrt{\lambda_o}}{L}\to 0$
(not shown).

In order to solve the problem of unphysical particles, we truncate the
tails of the distributions \emph{P(b-c)} for both radial and
cylindrical model, discarding the particles covering more than three
branch-cuts. This operation is relatively fast because we know at each
generation the list of the exposed branch-cut pre-images on the unit
circle (see (\ref{16})). Then, for each particle added, we calculate
how many branch-cuts of the previous generation the new particle
covers. If this number is $\le 3$ the particle is accepted, otherwise
it is discarded and a new attempt of particle growth is made.

As noticed in~\cite{21}, it is not clear \emph{a priori} whether the
growth which discards too large particles is macroscopically
equivalent to that without restrictions and whether such models
reproduce the same lattice aggregates. However for radial and
cylindrical cases the only significant proof is the comparison between
the scaling exponents derived numerically and the dimensions
accredited.

\noi For this reason we plot the quantities $F^{(n)}_1$ and $\ln
F^{(n)}_1$ in panel (c) and (d) of Fig.\ref{fig7}, related
respectively to the overall radius of a radial cluster
(\ref{5}) and to the height of a cylindrical DLA (see (\ref{19}) in
Section IV). Solid lines refer to models on which any particle
selection criterion is absent, whereas the dashed ones are related to
the proposed controlled growth procedure. We first note as the slopes
of both the curves remain unchanged by the presence or the absence of
too large particles: this leads to conclude that the method of
discarding particles covering more than three branch-cuts don't
compromise  the data extrapolation of the dimension. Moreover these
are in accordance with the accredited values present in literature: $1.71$ for
radial DLA and $1.67$ for cylindrical aggregates (see next section).
However we  stress that our method allows to filter out the
fluctuations due to the presence of filling particle that inevitably
affect the curves in panels (c) and (d) of Fig.\ref{fig7}; such
noisy behavior become more and more apparent in  the cylindrical case
as the ratio  $\frac{\sqrt{\lambda_0}}{L}$ tends to 1.

\section{Dimension, self-affinity and self-similarity}

As already noticed the first Laurent coefficient scales as the radius
$R_n$ of the aggregate~\cite{13,20} and it consequently provides an
useful tool to study the scaling behavior of the cluster in the radial
case. This property also applies to the cylindrical cluster in its
radial deformation, so that the overall size of the cluster grown in
the $z$-plane (Fig.\ref{fig5}(b)) is well characterized by
$F_1^{(n)}$. Moreover the radius in the $z$-plane is related to the
height ($y$) in the physical plane $z'$ by the second of (\ref{8}), so
that the logarithm of $F_1^{(n)}$ should represent the overall height
of the cluster. Under the hypothesis that the cylindrical DLA is
self-similar in squares of size $L/\sqrt{\lambda_0}$, the mean height
of a cylindrical cluster composed by $n$ particles growing in a stripe
of size $L/\sqrt{\lambda_0}$ scales as:

\be
y(n)\sim\frac{\sqrt{\lambda_0}^D}{L^{D-1}}n
\label{18}
\ee

\noi where $\sqrt{\lambda_0}$ is the linear size of the particles. We
can thus derive the scaling relation for the first Laurent coefficient
in the case of a cylindrical cluster:

\be
\ln F_1^{(n)}\sim 2\pi \left(\frac{\sqrt{\lambda_0}}{L}\right)^D n.
\label{19}
\ee

\noi It is important to remark that the self-similar growth (\ref{18})
of a cylindrical DLA is attained only in the steady state regime
~\cite{evertsz}, i.e. for $n\gg
\left(\frac{L}{\sqrt{\lambda_0}}\right)^D$.

\begin{figure}[!t]
\begin{center}
\includegraphics[width=8cm]{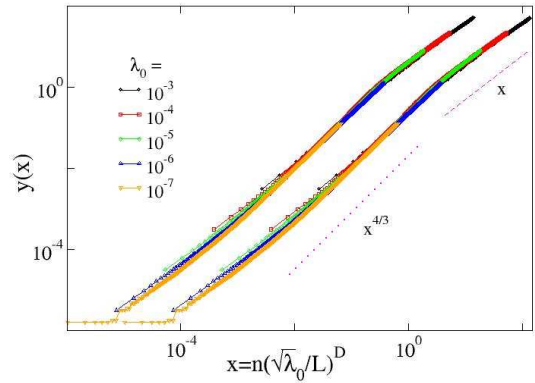}
\caption{Overall height of a cylindrical DLA obtained for different
  values of the particle size $\sqrt{\lambda_0}$; the collapse on the
  universal scaling function (\ref{affineconformal}) is apparent under
  the rescaling $x=n\left(\frac{\sqrt{\lambda_0}}{L}\right)^D$; the
  two different behaviours of $g(x)$ in (\ref{affineconformal}) are
  also displayed for comparison: $\sim x^{1.35\pm0.03}$ (dotted line),
  and $\sim x$ (dashed line). Statistics were taken over 20 different
  realizations with $L=1$. Two different collapse are displayed
  corresponding to $D=1.67$ and $D=1.71$ (shifted for clarity of one
  decade to the right). }
\label{fig8}
\end{center}
\end{figure}

Along the same lines followed in~\cite{24,25} for radial clusters, we
can argue that $\ln F_1^{(n)}\left(\frac{\sqrt{\lambda_0}}{L}\right)$
converges to a fixed point function $\left(\ln F_1\right)^{*}$ of the
single \emph{scaling} variable
$x=\left(\frac{\sqrt{\lambda_0}}{L}\right)^D n$, attaining the linear
regime (\ref{18}) asymptotically ($x\gg 1$). In Fig.\ref{fig8} we
present the average height $y(n)= \frac{L}{2 \pi} \ln
F_1^{(n)}\left(\frac{\sqrt{\lambda_0}}{L}\right)$ as a function of $x$
for a typical DLA realization for $L=1$ and $\sqrt{\lambda_0}$ ranging
from $10^{-3}$ to $10^{-7}$. Two different collapses are presented
setting $D=1.67$ and $D=1.71$.

It is evident how all the curves reasonably collapse on to a unique
scaling function
\be
y(x) = L g(x),
\label{affineconformal}
\ee

with

\be
g(x)\sim
\left\{
\begin{array}{ccc}
x^{1+\beta} &  & y\ll1\\
x &  & y\gg1
\end{array}
\right.
\label{affine_conformal_2}
\ee

\noi and $\beta=0.35 \pm 0.03$, obtained by averaging over the slopes
of different curves.

\noi Several remarks are in order:

\noi i) the convergence of $\ln F_1^{(n)}$ to the fixed point function
is obtained infinitesimally close to the cylinder base line, for which
$\ln F_1^{(n)}=0$;

\noi ii) the fixed point function exists already for $x\ll 1$: in this
transient regime the heigth of the aggregate exhibits the power-law
behavior $\sim x^{\alpha}$, with $\alpha \simeq \frac{4}{3}$.;

\noi iii) the predicted linear behavior is reached for $x\ge1$.

\noi The first two properties are also satisfied by radial
clusters~\cite{25}, while the third is a specific feature of
cylindrical aggregates.

The property i) refers to the stage of growth before the collapse and
can be roughly explained as follows.

\noi For cylindrical self-similar clusters, if we call $n_c$  the number
of particles required to obtain one-layer coverage of the original
circular interface, then $n_c\sim
\left(\frac{\sqrt{\lambda_0}}{L}\right)^{(1-D)}$. Therefore
$n_c\left(\frac{\sqrt{\lambda_0}}{L}\right)^D \to 0$ if
$\frac{\sqrt{\lambda_0}}{L}\to 0$.

The properties ii) and iii) refer to the fixed point function: in
particular ii) refers to the self-affine initial growth regime of the
aggregate. The matter of self-affine growth of cylindrical DLA has
been the subject of previous theoretical ~\cite{vannimenus} and
numerical investigations ~\cite{meakin,evertsz}, both showing that the
average cluster height grows as:

\be y(n)\sim L^{\frac{\nu_{||}}{\nu_\perp}}f\left(n\left
(\frac{\sqrt{\lambda_0}}{L}\right)^\frac{1}{\nu_\perp}\right),
\label{affinemeakin}
\ee

\noi where $\nu_{||}$ is the scaling exponent of a single DLA
\emph{tree} in the transverse direction, $\nu_{\perp}$ is the scaling
exponent in the growth direction and $f(x)$ is a scaling function that
behaves like $\sim x^{\frac{\nu_{||}}{1-\nu_\perp}}$ for $x\ll 1$
~\cite{Vic} and attains the linear regime for $x\gg 1$.

Our results indicate that the DLA dimension appears as a scaling
exponent even before the self-similar regime of growth (\ref{18}) sets
on, thus replacing both $\nu_{||}$ and $\nu_\perp$ in
(\ref{affinemeakin}). Note that setting $\nu_{||}=\frac{2}{3}$ and
$\nu_{\perp}=\frac{1}{2}$ in accordance with~\cite{meakin}, $g(x)$
coincides with $f(x)$ in (\ref{affinemeakin}).

A last remark concerns the estimate of the fractal dimension of the
cylindrical aggregate, as obtained with our conformal approach.  From
Fig.\ref{fig8} it is evident how the two data collapse are not
sufficiently sharp to discriminate between $D=1.67$ and $D=1.71$.
Additional numerical work (especially for
large sizes, i.e. very small values of $\sqrt{\lambda_0}$) would be
necessary to check whether the asymptotic fractal dimension of the
cylindrical DLA could be $D=1.71$, as in the radial case, as suggested
in~\cite{ball2}.

\section{Conclusions}

In this paper we examined carefully the numerical procedure to
generate the conformal maps in the cylindrical case. We have defined
the conformal map as the composition of two 
maps: one that maps the cylindrical aggregate in a radial one, and
another that map the radial aggregate in the circle. This procedure
opens up new prospects on the 
comprehension of the relationship between the aggregates growing on
different geometries. This also offers a conceptual advantage on the
definition of the ``right way'' to calculate the fractal dimension of
the DLA.

\noi Moreover we have proposed an auto-affine scaling relation for the
earlier stages of growth of the aggregate, on which the dimension of
the aggregate appears as scaling exponent. 
  
\ack It is a pleasure to acknowledge useful
discussion with Joachim Mathiensen about the numerical procedure used.

\end{document}